# Active Prostate Phantom with Multiple Chambers

Sizhe Tian[1], Yinoussa Adagolodjo[1], Jeremie Dequidt[1]

*Abstract*—Prostate cancer is a major global health concern, requiring advancements in robotic surgery and diagnostics to improve patient outcomes. A phantom is a specially designed object that simulates human tissues or organs. It can be used for calibrating and testing a medical process, as well as for training and research purposes. Existing prostate phantoms fail to simulate dynamic scenarios. This paper presents a pneumatically actuated prostate phantom with multiple independently controlled chambers, allowing for precise volumetric adjustments to replicate asymmetric and symmetric benign prostatic hyperplasia (BPH). The phantom is designed based on shape analysis of magnetic resonance imaging (MRI) datasets, modeled with finite element method (FEM), and validated through 3D reconstruction. The simulation results showed strong agreement with physical measurements, achieving average errors of 3.47% in forward modeling and 1.41% in inverse modeling. These results demonstrate the phantom's potential as a platform for validating robotic-assisted systems and for further development toward realistic simulation-based medical training.

## I. INTRODUCTION

Prostate cancer is one of the most common cancers in the world. Based on demographic data and rising life expectancy, a Lancet commission projected that the number of new prostate cancer cases annually would rise from 1.4 million in 2020 to 2.9 million by 2040 [1]. Early detection is crucial for improving patient outcomes.

Current screening techniques for prostate cancer include prostate-specific antigen (PSA) testing and digital rectal examination (DRE). PSA testing can be useful for evaluating possible prostate problems. However, it has a high rate of false positives in real-world clinical settings, which can lead to unnecessary biopsies [2]. A DRE involves a clinician palpating the prostate through the rectal wall to check for lumps or abnormal areas [3]. Combining DRE with PSA testing enhance the detection rates [4], [5]. However, many medical students feel unprepared to perform DRE due to a lack of hands-on training [6], which may contribute to missed or delayed diagnoses. A realistic replicate of prostate, or "phantom", could improve their training and lead to more accurate diagnoses. For cases requiring definitive diagnosis, prostate biopsies may become necessary [7]. Although effective, this invasive procedure can be uncomfortable for patients and technically challenging for surgeons. Robotic-assisted biopsy and surgical systems offers improved precision and minimized invasiveness, benefiting both patient comfort and surgical accuracy [8]. However, these systems require extensive testing and validation. Animal models, traditionally used for medical equipment development, raise ethical concerns, are expensive, and differ anatomically from human prostates. Therefore, realistic, cost-effective prostate phantoms are needed for training and system validation.

Existing commercially available phantoms [9] are overly simplistic. They are made of silicone blocks that fail to replicate realistic organ movement, volumetric changes, or mechanical responses. This limits the effectiveness of training for DRE and testing for robotic-assisted interventions. An advanced sensorized and actuated phantom capable of mimicking prostate deformation and responding to external stimuli would significantly enhance surgical training and robotic system validation [10], [11].

In this paper, we present a novel pneumatically actuated prostate phantom to address this gap. The phantom is designed to simulate the volumetric changes associated with benign prostatic hyperplasia (BPH) - a common condition characterized by prostate enlargement. We developed our phantom based on analysis of MRI dataset, implemented a finite element method (FEM) model to simulate its behaviour, then validate the numerical predictions against physical measurements. This approach allows for precise control and realistic simulation of prostate pathological change, crucial for both medical education and the testing of medical robotic systems. The rest of the paper is structured as follows: Section II reviews related work on medical and robotic phantoms. Section III details the development of our prostate phantom, including the shape analysis of the MRI data, the phantom design, modeling, and testing. Section IV presents the results of our work. Section V concludes the paper with a discussion of potential future improvements.

## II. RELATED WORK

Medical phantoms play a crucial role in developing and validating medical devices. Numerous researchers aim to build anatomically accurate, MRI and/or computed tomography (CT) compatible phantoms for various medical procedures [12]. For instance, Choi *et al.* developed a high-fidelity prostate phantom to simulate and evaluate transurethral resection of the prostate, demonstrating its potential in surgical training and procedure optimization [13]. Similarly, Betrouni *et al.* presented an anatomically realistic and adaptable prostate phantom for prostate cancer laser-based thermotherapy treatment planning and simulation [14]. Lindner *et al.* constructed anatomically correct models incorporating tumor, rectum, and

[1] Inria, CNRS, Centrale Lille, UMR 9189 CRIStAL, University of Lille, Lille, France.

Sizhe Tian is supported by both the ANR Region Hauts de France and the University of Lille (grant number ALRC 2.0-001143(P))

urethra that can be used for simulated and experimental magnetic resonance-guided focal intervention [15]. Researchers have developed sophisticated soft phantoms for many different organs. Öpik *et al.* designed kidney and liver phantoms with physical properties closely matching those of porcine organs, providing valuable tools for surgical training and device testing [16]. Additionally, Ghauri *et al.* presented a phantom model to mimic anatomical structures for validating dosimetry in prostate cancer photodynamic therapy [17]. Despite these advances, most existing phantoms are passive and lack sensing or actuation capabilities, limiting their uses in dynamic testing scenarios.

To address this limitation, recent research have sought to incorporate sensing and actuation abilities into phantoms. For example, Belcher *et al.* presented a 6-degree-of-freedom robotic phantom that can perform highly accurate motion along 6 axes to simulate the dynamic motion of a tumor for radiation therapy [18]. Shiinoki *et al.* developed a dynamic moving phantom system that can reproduce patient tumor motion and patient anatomy to improve the tracking accuracy of radiotherapy for lung cancer [19]. Additionally, He *et al.* presented an abdominal phantom for palpation training with controllable stiffness liver nodules that can also sense palpation forces [20]. Vogt *et al.* introduced an MRI-compatible abdomen phantom to simulate human organ movement and anatomy, aiding in medical training and device development [21]. Naghibi *et al.* developed an MRI-compatible soft robotic diaphragm phantom that is able to simulate liver respiratory motion using pneumatic actuation [22]. Ehrbar *et al.* presented ELPHA, a deformable liver phantom designed for testing motion detection and mitigation in real-time motion-adaptive radiotherapy [23]. Similarly, Hughes *et al.* introduce a sensorized soft robotic phantom designed to train and assess medical palpation techniques by providing feedback on the depth and location of palpation [24].

Specifically concerning prostate phantoms, research in integrating actuation and sensing capabilities remains limited. One example is the ADAM-pelvis phantom by Niebuhr *et al.*, a realistic, deformable male pelvis model that simulates human tissue and organ motion. It is MRI and CT compatible and can be used for validating image-guided radiotherapy [25]. Furthermore, Navarro *et al.* presented an MRI-safe prostate phantom with actuation and sensing ability, as well as a numerical simulation for the phantom. It can simulate the small displacements and volume growth due to inflammation and sense the force acting on the known location [26]. However, the volumetric transformations in these models were either limited in range or lacked fine control over asymmetric deformations.

Compared to previous works, our approach offers several contributions. We captured and characterized prostate shape variations using based on MRI data. Our phantom is real patient-derived, ensuring anatomical accuracy. Additionally, our design includes multiple independently controlled chambers, allowing for active, dynamic deformation to simulate both symmetric and asymmetric volumetric growth.

## III. METHODS

### A. Shape Analysis

Understanding shape variations is crucial, as prostate morphology varies significantly between individuals [27], [28]. This variation can influence results of prostate related procedures. To build a representative and anatomically accurate phantom, we used MRI data from the Cross-institution Male Pelvic Structures dataset [29]. This dataset contains 589 samples with labeled anatomical segmentations from seven different institutions.

To capture and analyse the prostate variation, we used Variational Autoencoder (VAE) [30]. A VAE is a generative neural network that learns a lower-dimensional latent space representation of the data. This compressed representation allows the VAE to effectively perform dimensionality reduction. The VAE consists of an encoder and a decoder. The encoder maps the input data to the latent space. And decoder generates models from that latent representation. By encoding the 3D prostate shapes into this latent space, we could analyze the shape differences and create an "average" prostate model which represents the most common anatomical features in the dataset.

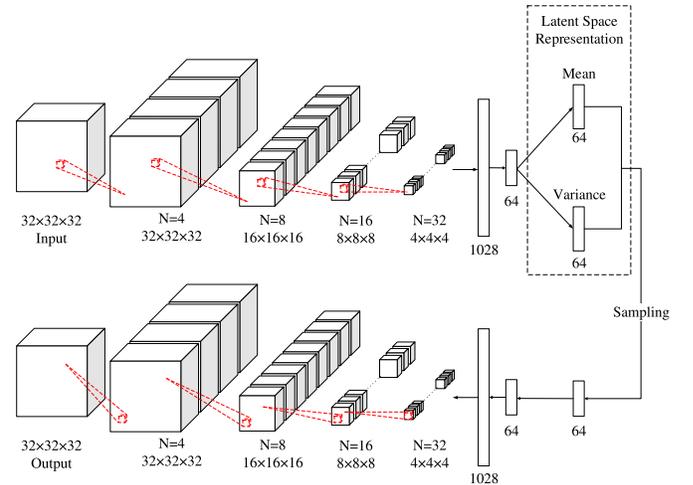

Fig. 1. Variational Autoencoder (VAE) architecture. The first row represents the encoder, and the second row represents the decoder. The dimensions of each layer are listed below. Each layer applies batch normalization and LeakyReLU activations.

*1) Model architecture:* We implemented a 3D convolutional VAE architecture, inspired by [31]. The network's size was reduced to avoid overfitting as the 3d structures of prostate model are relatively consistent. The 3D prostate models were voxelized to 32×32×32 resolution and feed into the VAE network. The architecture of the VAE network are shown in Figure 1. The latent space of VAE we used are 64-dimensional and the 3d convolution filter have a size of 3×3×3.

The loss function combined Binary Cross-Entropy (BCE) for reconstruction accuracy and Kullback-Leibler Divergence (KLD) for latent space regularization:

$$\mathcal{L} = \text{BCE}(\hat{x}, x) + \lambda \cdot \text{KLD}$$

where $x$ is input and the $\hat{x}$ is reconstructed output. $\lambda = 0.01$ balances reconstruction accuracy and latent space smoothness. The network was trained using a batch size of 10 for 500 epochs.

After training, all prostate models were encoded into the latent space, and their average latent representation was computed. The decoder then generated an "average prostate" from this representation.

*2) Model selection:* We chose a real patient-derived prostate model to ensure anatomical accuracy. While the VAE-generated model provided a statistically averaged prostate shape, the chosen patient-derived model ensures a closer representation of actual anatomical structures. The trained VAE network remains useful in evaluating the selected model's shape, analyzing prostate variations, and validating the phantom's design. The patient-derived prostate model was reconstructed using Screened Poisson surface reconstruction algorithm, smoothed, and scaled to match the average prostate volume [32], [33]. To verify similarity, we aligned the average model to the patient-derived model using Iterative Closest Point (ICP) alignment and computed the mean closest point distance between the two meshes [34]. This anatomical analysis ensures that our phantom is not only realistic but also adaptable for simulating a wide range of prostate deformations observed in clinical cases.

### B. System Design

*1) Phantom design:* The prostate may undergo different pathological changes. Two common pathological changes include BPH and carcinoma [35]. BPH, or prostate enlargement, is characterized by the growth of the prostate gland and may obstruct the prostatic urethra [35]. Carcinoma involves malignant tumor formation, which can lead to increased stiffness and tissue deformation [36]. In the scope of this paper, we focus on simulating BPH-related volumetric changes. Research indicates that men with symptomatic BPH exhibit a baseline prostate volume of 43.7 ± 0.38 ml, nearly double the normal volume of 26.3 ± 0.49 ml [33]. Traditional prostate phantoms are often static and fail to replicate asymmetric or progressive prostate growth. This type of growth is considered a suspicious finding during a DRE [4]. The multi-chamber design of our phantom allows for controlled, smooth, region-specific deformation patterns within a single model, instead of requiring multiple phantoms for different BPH patterns.

To accurately mimic BPH, we incorporated three independently controlled chambers in the prostate phantom: Chambers 1 and 2 are positioned on either side of the prostate gland to simulate asymmetrical enlargement. Chamber 0 is placed at the base of the prostate to allow for uniform volumetric expansion. This design is shown in Figure 2(a).

We fabricated the prostate phantom using lost-wax casting [37]. This method allowed us to create complex internal pneumatic chambers. The phantom was made of silicone gel[1], which has Young's modulus of approximately 50 kPa [38].

[1]Smooth-On EcoFlex 10

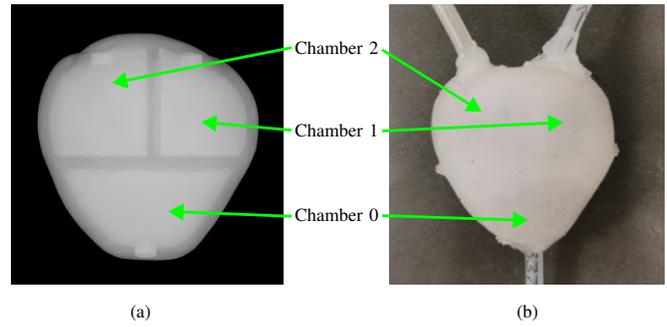

Fig. 2. Prostate model (a) Prostate model generated from MRI scans, showing three integrated actuation chambers. (b) Fabricated physical prostate phantom.

For reference, Carson *et al.* reported the elastic modulus of ex vivo prostate to be 61.8kPa ± 59.8kPa [39]. The material we chose is close to the average elastic modulus.

*2) Actuation:* We developed an actuation system to create controlled volumetric changes in the prostate phantom. To achieve this behaviour, we implemented a pneumatic actuation system. This system allows for controlled volumetric changes, smooth and continuous deformation, and is MRI-compatible. We selected Microblowers[2] as air pump to achieve precise volumetric expansion. The microblowers are selected over the conventional pump for their compact size, lightweight design, and rapid response time. The phantom contains three independently controlled chambers embedded within the phantom, each connected to a microblower. A closed-loop pressure control system regulates the inflation using integrated pressure sensors. Figure 3 illustrates the system architecture, showing the interaction between the controller, microblowers, pressure sensors [3], scanner, and the phantom.

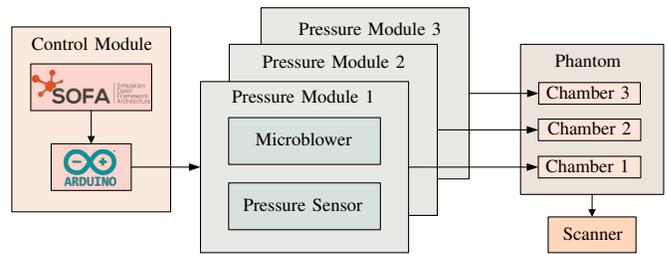

Fig. 3. Control system for the prostate phantom. The phantom includes three independently actuated chambers, each driven by microblowers. Pressure sensors provide real-time feedback, enabling pressure regulation by the controller. A 3D scanner captures the phantom's shape for reconstruction and evaluation. The control module combines a SOFA-based simulation with an Arduino hardware interface.

### C. Modeling

*1) Forward modeling:* We use FEM to model the prostate phantom numerically. The numerical simulation is implemented using Simulation Open Framework Architecture (SOFA) [40], with its SoftRobot and SoftRobot Inverse plugins

[2]Murata MZB3004T04
[3]Freescale Semiconductor MPXV7025G

[41]. We assumed the phantom's deformation during the inflation cycle was slow enough for a quasi-static simulation. This assumption is valid due to the low mass of the silicone model, which minimizes inertial effects and allows us to neglect dynamic forces. For simplicity, we also assume that the silicone behaves as a linear material, leading to elastic and reversible deformations. This approximation enables us to focus on the fundamental elastic behavior without introducing the complexities associated with nonlinear material properties. Based on these assumptions, we simulate the behavior of the prostate by solving a quasi-static co-rotational FEM model, as described in [42]. This approach captures the elastic response under deformation while accounting for rotational effects to improve accuracy, which yields good geometrical simulation results under large deformation [43]. The governing equation of the model is given by:

$$\mathbf{F(x_p)} - \mathbf{P(x_p)} - \mathbf{G} - \mathbf{H_p^T}\lambda = \mathbf{0} \quad (1)$$

In the following sections, we will break down each term of the governing equations.

*a) External constraints:* $\mathbf{G}$ corresponds to external constraints such as gravity.

*b) Internal forces:* $\mathbf{F(x_p)}$ represent the internal elastic forces. The variable $\mathbf{x_p}$ represents the displacements of the nodes in a tetrahedral mesh of the prostate model. $\mathbf{F(x_p)}$ is computed iteratively at each solver iteration $i$ using the following relation:

$$\mathbf{F(x_{p,i})} \approx \mathbf{R_e K_e (R_e^T x_p - x_0)} + \mathbf{F(x_{p,i-1})} \quad (2)$$

In this equation: $\mathbf{K_e}$ denotes the stiffness matrix of each tetrahedral element. It is calculated according to Hooke's law, which assumes linear elasticity. $\mathbf{R_e}$ is a block-diagonal matrix that includes four rotation matrices, each representing the rigid rotation of a tetrahedral element. These rotations account for element orientations, allowing the model to capture deformations more accurately under quasi-static assumptions. To improve stability and ensure convergence, the internal force vector $\mathbf{F(x_p)}$ is linearized between successive time steps using a first-order interpolation. This approach enables us to approximate the nonlinear behavior of the internal forces iteratively, thus improving computational efficiency while maintaining accuracy in capturing the prostate's elastic response. This iterative calculation of the internal forces allows the model to respond accurately to displacement changes over time, accommodating quasi-static deformations without introducing significant computational complexity.

*c) Fixation forces:* The fixation forces $\mathbf{P(x_p)}$ are applied to secure the prostate model in place. To fix the prostate model at the pneumatic actuation tube junction, we constrain the displacement of the associated nodes $X_{\text{fix}}$ to remain fixed in space. This is achieved using a penalization method, where each node in $X_{\text{fix}}$ is connected to its initial position $(\mathbf{x_{i,0}})$ at the start of the simulation through a virtual spring. This penalization approach ensures that nodes in $X_{\text{fix}}$ are effectively "locked" in place, while nodes outside this set $((i \notin X_{\text{fix}}))$ remain unconstrained, with no additional forces applied. By applying this high-stiffness constraint, the prostate model is fixed in space at the actuation tube junction, allowing the rest of the mesh to deform freely.

*d) Pressure actuation:* Pressure actuation of the prostate phantom is achieved through three internal cavities. Cavity walls are represented by surface meshes, linked to the 3D tetrahedral mesh via barycentric mapping. This bidirectional coupling ensures cavity wall deformation accurately reflects the 3D mesh, and forces on the walls propagate back to the 3D mesh. Pneumatic actuation applies pressure to the cavity surfaces, resulting in a force on node $i$ given by:

$$f_i = \sum_{t \in \mathcal{S}} \frac{a_t}{3} n_t \lambda_a \quad (3)$$

Where for each triangle $t$ of the chamber in the cavity triangles set $\mathcal{S}$, the product of its area $a_t$, its normal direction $n_t$, and pressure $\lambda_a$ yields the vector of force applied by the pneumatic actuation on the triangle $t$. This force is distributed equally among the triangle's three nodes. These forces are incorporated into the static model as constraints using Lagrange multipliers, represented by $H^T\lambda$, where $\lambda$ represents the force magnitudes. $H$ represents the gradient of these constraints with respect to the nodes in contact with the cavities. Specifically, $\lambda$ encodes the influence of pressure-induced forces on the nodes of the 3D mesh that define the cavity walls. This approach allows us to capture the effect of internal pressure on the deformation of the prostate model, translating cavity pressures into corresponding forces on the mesh nodes. The use of Lagrange multipliers facilitates enforcing these constraints in a manner that is computationally efficient while maintaining the fidelity of the model. Then we can solve the system under the assumption of a static equilibrium at each timestep, the equation (1).

*2) Inverse modeling:* For inverse modeling, we determine the required actuation to achieve a target deformation. More specifically, we aim to find the required pressure to reach a target volume. To solve this problem, we define a variable $\delta$ to evaluate the difference between the target volume and current volume, which can be written as:

$$\delta = \delta^{\text{free}} - W_{va}\lambda_a \quad (4)$$

Where $W_{va}$ is the compliance matrix that maps actuation force to volume change, $W_{va}\lambda_a$ represent the volume change incurred by the actuation $\lambda_a$, $\delta^{\text{free}}$ is the difference between the target volume and the volume of free configuration when no actuation is applied. Then, we can find the desired actuation by solving a Quadratic Programming (QP) problem:

$$\begin{aligned}
\lambda_a^* = \arg\min_{\lambda_a} \quad & \frac{1}{2}\delta^T\delta \\
\text{subject to} \quad & \delta_{\min} < \delta < \delta_{\max} \\
& \lambda_a \geq 0
\end{aligned} \quad (5)$$

The optimal solution $\lambda_a^*$ represents the desired pressure actuation. By applying the method described in [41], we can solve this QP problem to determine the pressure actuation that minimizes the difference between the target volume and the desired volume, thereby solving the inverse problem.

### D. Testing

We evaluate the performance of our phantom through experimental validation and compare the numerical model and physical phantom. We measure the physical phantom by reconstructing its 3D model using Meshroom [44]. To collect the image needed for 3D reconstruction, we built a scanner shown in Figure 4. The scanner takes more than 100 pictures around the actuated physical phantom. Then we run the pipeline in Meshroom to get the 3D model of the physical phantom for further analysis.

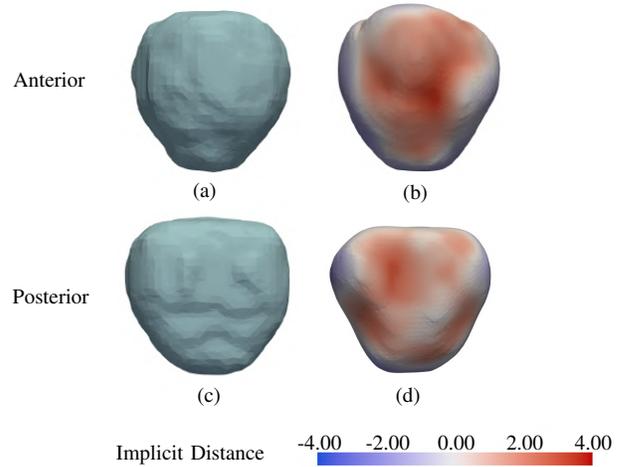

Fig. 5. (a)(c)Anterior and posterior views of the VAE-generated average prostate model (smoothed). (b)(d) Comparison of the average model and the selected patient-derived prostate model for the phantom, visualized as an implicit distance heatmap on the chosen model.

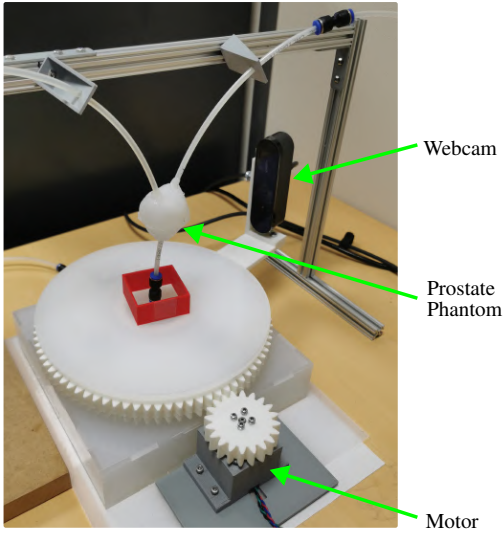

Fig. 4. Setup for 3D reconstruction of the prostate phantom. The phantom is suspended in air, while a webcam mounted on a motorized platform captures multiple images for volumetric assessment.

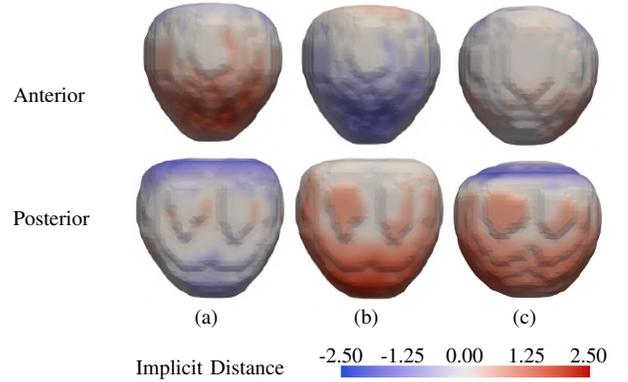

Fig. 6. Comparison of three cluster centers in the latent space, reconstructed using the decoder. The implicit distance heatmap on the selected model visualizes shape differences: (a) Cluster 1 vs. Cluster 2, (b) Cluster 1 vs. Cluster 3, (c) Cluster 2 vs. Cluster 3.

## IV. RESULTS

### A. Shape variation

As presented in Section III-A, We generated the average prostate model from VAE. Figure 5 compares the VAE-generated model with the selected patient-derived model used for the phantom. The volume of the VAE-generated prostate was 27.3 ml, while the selected model was 26.3 ml. The mean closest point distance between the two aligned models was 1.00 mm, and their vector distance in latent space was 0.64. These results indicate the chosen model closely resembles the statistical average prostate shape.

To further analyze shape variations, we clustered the latent space representations using the k-means algorithm and reconstructed each cluster center using the decoder. Figure 6 shows comparisons between each cluster center model as implicit distance heatmaps [34]. The reconstruction of cluster centers revealed that most shape variations occurred along the anteroposterior axis and proximodistal axis, which agrees with the observations reported in [27]. Notably, significant variations were observed in the lower region of the prostate, justifying the inclusion of a third chamber at the phantom's base.

### B. Forward Model

We evaluated the accuracy of the forward model by comparing simulated and experimental results. Figure 7 showcases the phantom in its initial state and post-inflation state, comparing simulated, physical, and reconstructed models. A visible ring is presented on the physical phantom due to the casting process. However, this artifact had negligible influence on volumetric expansion and was not considered in the subsequent analysis. Empirical characterization of the forward model was performed through actual volume and simulation comparisons. The Young's modulus was subsequently manually tuned using only few critical data points from forward simulation results presented, yielding an estimated value of 1.65 MPa. This esti-mation differs from previously mentioned material properties due to factors such as meshing resolution, modeling inaccuracy and simplifications inherent in the model. The results are listed in Table I, where $P_0$, $P_1$, and $P_2$ denote pressure inputs as referenced in Figure 2, $V_M$ and $V_S$ are the measured and simulated error respectively, $e$ represent errors. The measured and simulated volumes closely matched, with an average error of 3.47%. However, as volume increased, error rates also increased, which could be a result of the non-linearity during large elastic deformation.

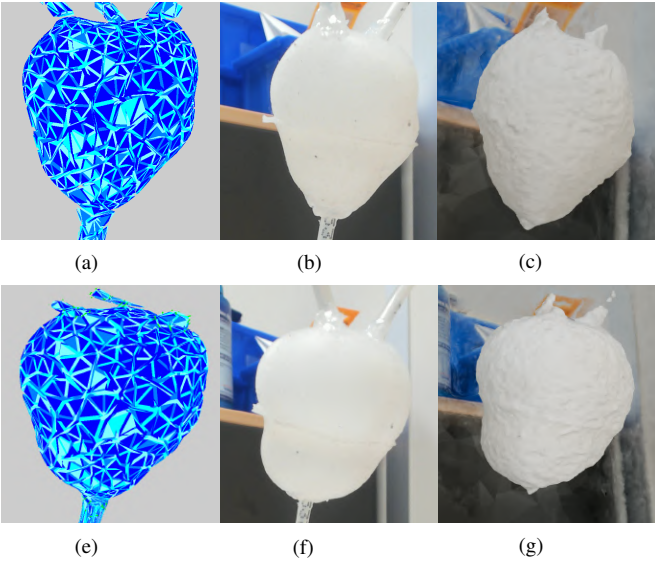

Fig. 7. Comparison of the phantom in simulation, physical world, and reconstruction. (a) Initial state in simulation, (b) Initial state of the physical phantom, (c) 3D reconstruction of the physical phantom in the initial state, (d) Simulation of the inflated phantom, (f) Inflated physical phantom (g) 3D reconstruction of the inflated physical model

TABLE I
RESULT OF THE FORWARD SIMULATION

| $P_0$ (kPa) | $P_1$ (kPa) | $P_2$ (kPa) | $V_M$ (ml) | $V_S$ (ml) | $e$ (ml) |
|---|---|---|---|---|---|
| 0 | 0 | 0 | 26.43 | 26.33 | 0.1 |
| 1 | 1 | 1 | 29.96 | 30.00 | -0.04 |
| 2 | 2 | 2 | 34.80 | 35.90 | -1.1 |
| 2 | 0 | 0 | 32.36 | 29.40 | 2.96 |
| 0 | 2 | 0 | 32.17 | 29.00 | 3.17 |
| 0 | 0 | 2 | 31.20 | 29.47 | 1.73 |
| 1 | 2 | 3 | 38.54 | 39.14 | -0.6 |
| 3 | 1 | 2 | 40.10 | 38.10 | 2 |

### C. Inverse Model

For the inverse problem, we specified a target volume and solved for the necessary pressure to achieve that volume using the estimated Young's modulus from the forward model. The results, illustrated in Figure 8, show that the measured volumes closely aligned with target volumes, with an average error of 1.41%. The inverse model demonstrated slightly higher accuracy than the forward model. This may be because the optimization process distributes pressures uniformly across the chambers. In contrast, the forward model's direct pressure application may create unintended pressure variations due to chamber interactions and the unidirectional nature of the microblower pump.

To further evaluate the numeric model, we compare the reconstructed phantom shapes with the simulated phantom shape with different target volume. Figure 9 compares the mean closest point distance and distance in latent space between two models. The latent space distance is computed by encoding the 3D models using the previously trained VAE to obtain its latent vector, and then calculating the Euclidean distance between the two. While the shapes are generally similar, discrepancies appear with increasing target volume. This is further illustrated in Figure 10, which presents a heatmap of the implicit distance between the two models. As the target volume increases, so does the discrepancy, primarily along the anteroposterior axis. Additionally, Figure 10 suggests that the physical phantom tends to be rounded under higher actuation and the simulation does not fully capture this change.

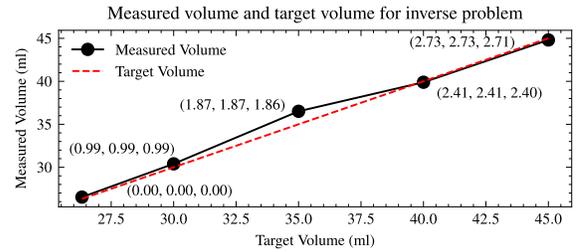

Fig. 8. Comparison of measured and target volumes in the inverse model. The required actuation pressures ($P_0$, $P_1$, $P_2$, in kPa) for each target volume are labeled next to the data points.

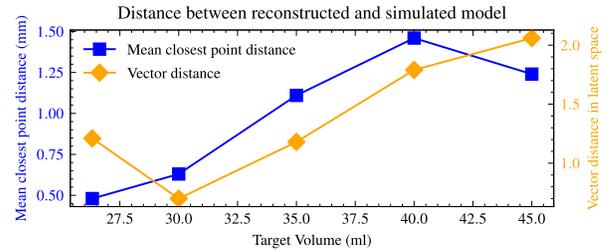

Fig. 9. Comparison of reconstructed and simulated prostate phantom models. Mean closest point distance and latent space vector distance are used to measure shape deviations.

### D. Summary

The evaluation of both the forward and inverse models confirms that the proposed prostate phantom effectively simulates the volumetric changes associated with BPH. The forward model demonstrated an average error of 3.47%, with increased discrepancies at higher deformations possibly due to material nonlinearity. The inverse model achieved better accuracy (1.41%), likely due to the pressure optimization process. Despite the similarity between simulations and physical measurements, some discrepancies are observed, especially at higher pressure. This suggests that our FEM model could use some further improvement like a refined material model to account for non-linearity, or better meshing strategy to capture deformation pattern more accurately. Overall, these results validate the multi-chamber pneumatic actuation approach, demonstrating its potential for robotic system validation and medical training.

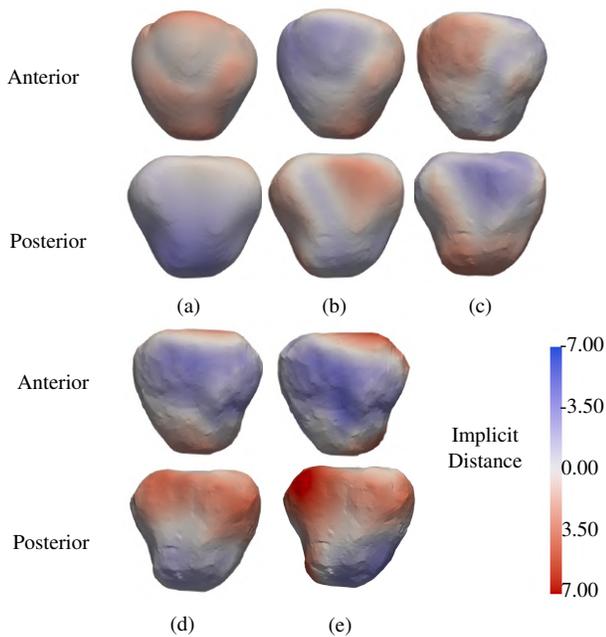

Fig. 10. Comparison of the reconstructed and simulated prostate phantom models. Implicit distance heatmaps on simulated models visualize differences between the two models for target volumes of: (a) 26 ml, (b) 30 ml, (c) 35 ml, (d) 40 ml, and (e) 45 ml.

## V. Conclusion

This paper presented an pneumatically actuated prostate phantom designed to simulate BPH by mimicking volumetric changes. By integrating multiple independently controlled chambers, our design allows for asymmetric and symmetric expansion, improving anatomical realism for both medical training and robotic system validation. The phantom was developed based on VAE-based shape analysis on an MRI dataset and validated through both numerical simulations and physical experiments. Our results demonstrated strong agreement between the simulated and reconstructed models, confirming the model's reliability in replicating volumetric prostate changes. The system offers a potential tool for both medical training and robotic system validation by providing a more diverse and realistic simulation of the prostate.

The validation of the numerical model and the physical phantom demonstrated high accuracy. However, there are still some limitations to address. First, while our phantom successfully replicates volumetric changes, it lacks interactive feedback for external manipulation. This limits its application for evaluating digital rectal examination (DRE) techniques and robotic-assisted interventions that require real-time force sensing. Future work will focus on integrating embedded sensors to provide haptic feedback and enhance the training experience. Secondly, our phantom is currently limited to simulating BPH. The lack of variable stiffness and finer structural differentiation restricts its ability to model conditions such as prostate carcinoma. Integrating granular jamming or hybrid soft-rigid structures could enable future versions of the phantom to better replicate pathological variations in prostate tissue. Thirdly, we used silicone to mimic prostate tissue and assumed its linear elasticity for simulation. It is a simplified material model for a real prostate tissues, which could cause discrepancy. Future iterations could explore other materials model to improve biomechanical accuracy. Finally, the presence of a ring artifact due to the casting process slightly restricts the expansion of the phantom, and future iterations could explore more advanced manufacturing methods, such as 3D printing, to eliminate this issue.

Despite these limitations, our multi-chamber prostate phantom represents a step toward more physiologically relevant and adaptable models for medical education and robotic validation. By addressing these limitations, future developments can further improve its clinical applicability and expand its potential for training, diagnostics, and surgical planning.

## References

[1] N. D. James, I. Tannock, J. N'Dow, F. Feng, S. Gillessen, S. A. Ali, B. Trujillo, B. Al-Lazikani, G. Attard, F. Bray, and others, "The lancet commission on prostate cancer: planning for the surge in cases," vol. 403, no. 10437, pp. 1683–1722, publisher: Elsevier.

[2] B. Lumbreras, L. A. Parker, J. P. Caballero-Romeu, L. Gómez-Pérez, M. Puig-García, M. López-Garrigós, N. García, and I. Hernández-Aguado, "Variables associated with false-positive PSA results: A cohort study with real-world data," vol. 15, no. 1. [Online]. Available: https://www.mdpi.com/2072-6694/15/1/261

[3] "Prostate model - enlarged right lobe — gtsimulators.com." [Online]. Available: https://www.gtsimulators.com/products/prostate-model-enlarged-right-lobe-gp3000

[4] W. J. Catalona, J. P. Richie, F. R. Ahmann, M. A. Hudson, P. T. Scardino, R. C. Flanigan, J. B. DeKernion, T. L. Ratliff, L. R. Kavoussi, B. L. Dalkin, W. B. Waters, M. T. MacFarlane, and P. C. Southwick, "Comparison of digital rectal examination and serum prostate specific antigen in the early detection of prostate cancer: results of a multicenter clinical trial of 6,630 men." vol. 151, no. 5, pp. 1283–1290, place: United States.

[5] G. F. Carvalhal, D. S. Smith, D. E. Mager, C. Ramos, and W. J. Catalona, "Digital rectal examination for detecting prostate cancer at prostate specific antigen levels of 4 ng./ml. or less," vol. 161, no. 3, pp. 835–839. [Online]. Available: https://www.sciencedirect.com/science/article/pii/S0022534701617853

[6] C. Nikendei, K. Diefenbacher, N. Köhl-Hackert, H. Lauber, J. Huber, A. Herrmann-Werner, W. Herzog, J.-H. Schultz, J. Jünger, and M. Krautter, "Digital rectal examination skills: first training experiences, the motives and attitudes of standardized patients," vol. 15, no. 1, p. 7. [Online]. Available: https://doi.org/10.1186/s12909-015-0292-7

[7] "Prostate cancer treatment — cancer.gov." [Online]. Available: https://www.cancer.gov/types/prostate/patient/prostate-treatment-pdq

[8] S. S. Dhaliwal, A. Belarouci, M. S. Lopez, F. Verbrugghe, O. Lakhal, G. Dherbomez, T. Chettibi, and R. Merzouki, "CoBra: Towards adaptive robotized prostate brachytherapy under MRI guidance," in *2021 Sixth International Conference on Advances in Biomedical Engineering (ICABME)*, pp. 81–84, ISSN: 2377-5696.

[9] "Male rectal examination trainer - advanced (light skin tone) | limbs & things UK — limbsandthings.com." [Online]. Available: https://limbsandthings.com/uk/products/60171/60171-male-rectal-examination-trainer-advanced-light-skin-tone

[10] D. Zrinscak, L. Lorenzon, M. Maselli, and M. Cianchetti, "Soft robotics for physical simulators, artificial organs and implantable assistive devices," vol. 5, no. 1, p. 012002, publisher: IOP Publishing. [Online]. Available: https://dx.doi.org/10.1088/2516-1091/acb57a

[11] K. Gilday, I. Zubak, A. Raabe, and J. Hughes, "From rigid to soft robotic approaches for minimally invasive neurosurgery."

[12] S. Wilby, A. Palmer, W. Polak, and A. Bucchi, "A review of brachytherapy physical phantoms developed over the last 20 years: clinical purpose and future requirements," vol. 13, no. 1, pp. 101–115. [Online]. Available: http://dx.doi.org/10.5114/jcb.2021.103593

[13] E. Choi, F. Adams, S. Palagi, A. Gengenbacher, D. Schlager, P.-F. Müller, C. Gratzke, A. Miernik, P. Fischer, and T. Qiu, "A high-fidelity phantom for the simulation and quantitative evaluation of transurethral resection of the prostate," vol. 48, no. 1, pp. 437–446. [Online]. Available: https://doi.org/10.1007/s10439-019-02361-7

[14] N. Betrouni, P. Nevoux, B. Leroux, P. Colin, P. Puech, and S. Mordon, "An anatomically realistic and adaptable prostate phantom for laser thermotherapy treatment planning," vol. 40, no. 2, p. 022701, publisher: Wiley Online Library.

[15] U. Lindner, N. Lawrentschuk, R. A. Weersink, O. Raz, E. Hlasny, M. S. Sussman, S. R. Davidson, M. R. Gertner, and J. Trachtenberg, "Construction and evaluation of an anatomically correct multi-image modality compatible phantom for prostate cancer focal ablation," vol. 184, no. 1, pp. 352–357, publisher: Elsevier.

[16] R. Öpik, A. Hunt, A. Ristolainen, P. M. Aubin, and M. Kruusmaa, "Development of high fidelity liver and kidney phantom organs for use with robotic surgical systems," in *2012 4th IEEE RAS & EMBS International Conference on Biomedical Robotics and Biomechatronics (BioRob)*, pp. 425–430, ISSN: 2155-1782.

[17] M. D. Ghauri, N. Guadagno, S. Bhattacharya, B. Thomasson, J. Swartling, R. Gautam, S. Andersson-Engels, and S. K. V. Sekar, "Hybrid heterogeneous phantoms for biomedical applications: a demonstration to dosimetry validation," vol. 15, no. 2, pp. 863–874, publisher: Optica Publishing Group. [Online]. Available: https://opg.optica.org/boe/abstract.cfm?URI=boe-15-2-863

[18] A. H. Belcher, X. Liu, Z. Grelewicz, E. Pearson, and R. D. Wiersma, "Development of a 6dof robotic motion phantom for radiation therapy." vol. 41, no. 12, p. 121704, place: United States.

[19] T. Shiinoki, F. Fujii, K. Fujimoto, Y. Yuasa, and T. Sera, "A novel dynamic robotic moving phantom system for patient-specific quality assurance in real-time tumor-tracking radiotherapy." vol. 21, no. 7, pp. 16–28, place: United States.

[20] L. He, N. Herzig, S. d. Lusignan, L. Scimeca, P. Maiolino, F. Iida, and T. Nanayakkara, "An abdominal phantom with tunable stiffness nodules and force sensing capability for palpation training," vol. 37, no. 4, pp. 1051–1064.

[21] I. Vogt, K. Engel, A. Schlünz, R. Kowal, B. Hensen, M. Gutberlet, F. Wacker, and G. Rose, "MRI-compatible abdomen phantom to mimic respiratory-triggered organ movement while performing needle-based interventions." [Online]. Available: https://doi.org/10.1007/s11548-024-03188-x

[22] H. Naghibi, J. v. Dorp, and M. Abayazid, "Development of a soft robotics diaphragm to simulate respiratory motion," in *2020 8th IEEE RAS/EMBS International Conference for Biomedical Robotics and Biomechatronics (BioRob)*, pp. 140–145, ISSN: 2155-1782.

[23] S. Ehrbar, A. Jöhl, M. Kühni, M. Meboldt, E. Ozkan Elsen, C. Tanner, O. Goksel, S. Klöck, J. Unkelbach, M. Guckenberger, and S. Tanadini-Lang, "ELPHA: Dynamically deformable liver phantom for real-time motion-adaptive radiotherapy treatments," vol. 46, no. 2, pp. 839–850, _eprint: https://aapm.onlinelibrary.wiley.com/doi/pdf/10.1002/mp.13359. [Online]. Available: https://aapm.onlinelibrary.wiley.com/doi/abs/10.1002/mp.13359

[24] J. Hughes, P. Maiolino, T. Nanayakkara, and F. Iida, "Sensorized phantom for characterizing large area deformation of soft bodies for medical applications," in *2020 3rd IEEE International Conference on Soft Robotics (RoboSoft)*, pp. 278–284.

[25] N. I. Niebuhr, W. Johnen, G. Echner, A. Runz, M. Bach, M. Stoll, K. Giske, S. Greilich, and A. Pfaffenberger, "The ADAM-pelvis phantom—an anthropomorphic, deformable and multimodal phantom for MRgRT," vol. 64, no. 4, p. 04NT05, publisher: IOP Publishing. [Online]. Available: https://dx.doi.org/10.1088/1361-6560/aafd5f

[26] S. Escaida Navarro, S. S. Dhaliwal, M. S. Lopez, S. Wilby, A. L. Palmer, W. Polak, R. Merzouki, and C. Duriez, "A bio-inspired active prostate phantom for adaptive interventions," vol. 4, no. 2, pp. 300–310.

[27] M. Rusu, A. S. Purysko, S. Verma, J. Kiechle, J. Gollamudi, S. Ghose, K. Herrmann, V. Gulani, R. Paspulati, L. Ponsky, M. Böhm, A.-M. Haynes, D. Moses, R. Shnier, W. Delprado, J. Thompson, P. Stricker, and A. Madabhushi, "Computational imaging reveals shape differences between normal and malignant prostates on MRI," vol. 7, no. 1, p. 41261. [Online]. Available: https://doi.org/10.1038/srep41261

[28] S. Qian, X. Sheng, D. Xu, H. Shen, J. Qi, and Y. Wu, "Variation of prostatic morphology in chinese benign prostatic hyperplasia patients of different age decades," vol. 23, no. 5, pp. 457–463, publisher: Taylor & Francis. [Online]. Available: https://doi.org/10.1080/13685538.2018.1522626

[29] Y. Li, Y. Fu, I. J. M. B. Gayo, Q. Yang, Z. Min, S. U. Saeed, W. Yan, Y. Wang, J. A. Noble, M. Emberton, M. J. Clarkson, H. Huisman, D. C. Barratt, V. A. Prisacariu, and Y. Hu, "Prototypical few-shot segmentation for cross-institution male pelvic structures with spatial registration," vol. 90, p. 102935. [Online]. Available: https://www.sciencedirect.com/science/article/pii/S1361841523001950

[30] D. P. Kingma and M. Welling, "Auto-encoding variational bayes," _eprint: 1312.6114. [Online]. Available: https://arxiv.org/abs/1312.6114

[31] A. Brock, T. Lim, J. M. Ritchie, and N. Weston, "Generative and discriminative voxel modeling with convolutional neural networks."

[32] A. Muntoni and P. Cignoni, "PyMeshLab," Jan. 2021.

[33] C. G. Roehrborn, P. Boyle, A. L. Gould, and J. Waldstreicher, "Serum prostate-specific antigen as a predictor of prostate volume in men with benign prostatic hyperplasia," vol. 53, no. 3, pp. 581–589. [Online]. Available: https://www.sciencedirect.com/science/article/pii/S0090429598006554

[34] B. Sullivan and A. Kaszynski, "PyVista: 3d plotting and mesh analysis through a streamlined interface for the visualization toolkit (VTK)," vol. 4, no. 37, p. 1450, publisher: The Open Journal. [Online]. Available: https://doi.org/10.21105/joss.01450

[35] R. Cannarella, R. A. Condorelli, F. Barbagallo, S. La Vignera, and A. E. Calogero, "Endocrinology of the aging prostate: Current concepts." vol. 12, p. 554078, place: Switzerland.

[36] S. Ishihara and H. Haga, "Matrix stiffness contributes to cancer progression by regulating transcription factors." vol. 14, no. 4, place: Switzerland.

[37] A. D. Marchese, R. K. Katzschmann, and D. Rus, "A recipe for soft fluidic elastomer robots," vol. 2, no. 1, pp. 7–25, _eprint: https://doi.org/10.1089/soro.2014.0022. [Online]. Available: https://doi.org/10.1089/soro.2014.0022

[38] J. Vaicekauskaite, P. Mazurek, S. Vudayagiri, and A. L. Skov, "Mapping the mechanical and electrical properties of commercial silicone elastomer formulations for stretchable transducers," *Journal of Materials Chemistry C*, vol. 8, no. 4, pp. 1273–1279, 2020.

[39] W. C. Carson, G. J. Gerling, T. L. Krupski, C. G. Kowalik, J. C. Harper, and C. A. Moskaluk, "Material characterization of ex vivo prostate tissue via spherical indentation in the clinic," vol. 33, no. 3, pp. 302–309. [Online]. Available: https://www.sciencedirect.com/science/article/pii/S1350453310002444

[40] F. Faure, C. Duriez, H. Delingette, J. Allard, B. Gilles, S. Marchesseau, H. Talbot, H. Courtecuisse, G. Bousquet, I. Peterlik, and S. Cotin, "SOFA: A multi-model framework for interactive physical simulation," in *Soft Tissue Biomechanical Modeling for Computer Assisted Surgery*, ser. Studies in Mechanobiology, Tissue Engineering and Biomaterials, Y. Payan, Ed. Springer, vol. 11, pp. 283–321. [Online]. Available: https://inria.hal.science/hal-00681539

[41] E. Coevoet, T. Morales Bieze, F. Largilliere, Z. Zhang, M. Thieffry, M. Sanz Lopez, B. Carrez, D. Marchal, O. Goury, J. Dequidt, and C. Duriez, "Software toolkit for modeling, simulation and control of soft robots," vol. 31, pp. 1208–1224, publisher: Taylor & Francis. [Online]. Available: https://inria.hal.science/hal-01649355

[42] C. A. Felippa, "A systematic approach to the element-independent corotational dynamics of finite elements," publisher: Technical Report CU-CAS-00-03, Center for Aerospace Structures.

[43] M. Marchal, J. Allard, C. Duriez, and S. Cotin, "Towards a framework for assessing deformable models in medical simulation," in *Biomedical Simulation*, F. Bello and P. J. E. Edwards, Eds. Springer Berlin Heidelberg, pp. 176–184.

[44] C. Griwodz, S. Gasparini, L. Calvet, P. Gurdjos, F. Castan, B. Maujean, G. D. Lillo, and Y. Lanthony, "AliceVision meshroom: An open-source 3d reconstruction pipeline," in *Proceedings of the 12th ACM Multimedia Systems Conference - MMSys '21*. ACM Press.